\documentclass[submission,copyright,creativecommons]{eptcs}
\providecommand{\event}{PACO 2011}
\usepackage{breakurl}

\usepackage[english]{babel}
\usepackage[latin1]{inputenc}
\usepackage[T1]{fontenc}
\usepackage{epsfig}
\usepackage{graphicx}
\usepackage{amssymb}
\usepackage{amsfonts}
\usepackage{amsmath}
\usepackage{verbatim}
\usepackage{url}
\usepackage{xcolor}
\usepackage{gastex}
\usepackage{floatflt}
\usepackage{stmaryrd}
\usepackage[all]{xypic}
\usepackage{epic,eepic}
\usepackage{color}

\xdefinecolor{Red}{rgb}{1,0.2,0}
\xdefinecolor{Green}{rgb}{0,0.7,0}
\xdefinecolor{Yellow}{rgb}{1,1,0}
\xdefinecolor{White}{rgb}{1,1,1}
\xdefinecolor{Blue}{rgb}{0,0,1}
\xdefinecolor{Black}{rgb}{0,0,0}
\xdefinecolor{Orange}{rgb}{1,1,0}
\newtheorem{adefinizione}{Definition}[section]
\newtheorem{fatto}[adefinizione]{Fact}
\newtheorem{alemma}[adefinizione]{Lemma}
\newtheorem{teorema}[adefinizione]{Theorem}
\newtheorem{corollario}[adefinizione]{Corollary}
\newtheorem{proposizione}[adefinizione]{Proposition}
\newtheorem{esempio}[adefinizione]{Example}

\newenvironment{definition}{\begin{adefinizione}\ \rm}{\end{adefinizione}}

\title{Parameterized Verification of Safety Properties in\\
Ad Hoc Network Protocols}
\author{Giorgio Delzanno
\institute{University of Genova - Italy}
\email{delzanno@disi.unige.it}
\and
Arnaud Sangnier
\institute{LIAFA, University  Paris 7 - France}
\email{sangnier@liafa.jussieu.fr}
\and
Gianluigi Zavattaro
\institute{University of Bologna - Italy}
\email{zavattar@cs.unibo.it}
}
\def\titlerunning{Parameterized Verification of Safety Properties in
Ad Hoc Networks}
\def\authorrunning{G. Delzanno, A. Sangnier, and G. Zavattaro}
\newcommand{\set}[1]{\{ #1 \} }
\newcommand{\tuple}[1]{\langle #1 \rangle}

\newcommand{\broadcast}[1]{!!{#1}} 
\newcommand{\receive}[1]{??{#1}}

\newcommand{\inclique}[3]{{#1}\sim_{#3}{#2}}

\newcommand{\state}{L}

\newcommand{\confs}{{\cal C}}

\newcommand{\CSRP}{{\sc cover}}

\newcommand{\Topology}{{\cal T}}
\newcommand{\ahn}{ {\cal A} }
\newcommand{\pdef}{ {\cal P} }
\newcommand{\mytau}{\tau}
\newcommand{\ignore}[1]{}

\newif\iflong
\longfalse

\begin{document}
\maketitle
\begin{abstract}
We summarize the main results proved in recent work
on the parameterized verification of safety properties
for ad hoc network protocols. We consider a model in which
the communication topology of a network is represented as a graph. Nodes represent states of individual processes. Adjacent nodes represent single-hop neighbors. Processes are finite state automata that communicate via selective broadcast messages. Reception of a broadcast is restricted to single-hop neighbors. For this model we consider a decision problem that can be expressed as the verification of the existence of an initial topology in which the execution of the protocol can lead to
a configuration with at least one node in a certain state.
The decision problem is parametric both on the size and on the form of the communication topology of the initial configurations. We draw a complete picture of the decidability and complexity boundaries of this problem according to various assumptions on the possible topologies.
\end{abstract}
\section{Introduction}
Ad hoc networks consist of wireless hosts that, in the absence of a fixed infrastructure, communicate sending broadcast messages.
In this context protocols are typically supposed to work independently from the communication topology and from the size (number of nodes) of the network.
%
As suggested in \cite{delzanno-parameterized-10,delzanno-fossacs11}, the {\em control state reachability problem} (or {\em coverability problem}) seems a particularly adequate formalization  of parameterized verification problems for ad hoc networks.
A network is represented as a graph in which nodes are individual processes and edges represent communication links.
Each node executes an instance of the same protocol. A protocol is described by a finite state communicating automaton. 
The control state reachability problem consists in checking whether there exists an initial graph (with unknown size and topology) that can evolve into a configuration in which at least one node is in a given error state. Since the size of the initial configuration is not 
fixed a priori, the state-space to be explored is in general infinite. 

In this paper we summarize the main results that we have proved in 
two recent publications \cite{delzanno-parameterized-10,delzanno-fossacs11}. The first result is negative: control state reachability is undecidable if we do not fix any restriction on the possible topologies.
As for other communication models \cite{MeyerTCS08,Fossacs10}, finding interesting classes of network topologies  
for which verification is, at least theoretically, possible is an important research problem.
As a first positive result, we have proved in~\cite{delzanno-parameterized-10} that control state reachability turns out
to be decidable for the class of {\em bounded path graphs}. Graphs have bounded path if there exists a value $k$ such that
all simple paths in the considered graph have length smaller than $k$. 
Although for a fixed $k$ this class of graphs is infinite, it appears of limited
interest as it does not include clique graphs. 
Cliques are appealing for at least two reasons. First, they represent the best possible scenario for optimizing broadcast communication
(one broadcast to reach all nodes). 
Second, when restricting configurations only to cliques, 
control state reachability can be reduced to coverability in a Broadcast Protocol,
i.e., in a model in which configurations are multisets of processes defined by communicating automata \cite{Emerson-Namjoshi98}.
Coverability is decidable in Broadcast Protocols in \cite{Esparza-Finkel-Mayr-Lics99}.

For these reasons, in \cite{delzanno-fossacs11} we have decided
to investigate classes of graphs that at least include the clique
graphs.
More precisely,  we have considered networks in which the underlying topology  is in between the class of
 {\em cliques} and the strictly larger class of {\em bounded diameter graphs}.
Graphs have bounded diameter if there exists a value $k$ such that the minimal path between every pair of nodes of the same graph 
has length smaller than $k$. 
 Graphs with bounded diameter (also called clusters) are particularly relevant for the domain of ad hoc networks. They are often used to partition a network in order to increase the efficiency of broadcast communication \cite{Clusters}.

The restriction to bounded diameter 
follows the approach taken for point-to-point communication in 
\cite{MeyerTCS08,Fossacs10}.
Differently from \cite{MeyerTCS08,Fossacs10} we have proved
that for our model of selective broadcast
control state reachability is undecidable when restricting the topologies to graphs whose diameter is bounded by $k$  (for a fixed $k >0$).
Then, we have investigated further restrictions having in mind the constraint that they must allow at least cliques of arbitrary order.
By using an original well-quasi ordering result, we have proved that control state reachability becomes decidable when considering a class of graphs in which the corresponding maximal cliques are connected by paths of bounded length.
Furthermore, by exploiting a recent result of Schnoebelen \cite{Schnoebelen2010} and a reduction to coverability in reset nets, we 
have shown that the resulting decision procedure is Ackermann-hard. Interestingly, the same complexity result already holds in the subclass of clique topologies.
%

\paragraph{Related Work}
%
%
Ethernet-like broadcast communication has been analyzed by Prasad
\cite{Pra95} using the Calculus of Broadcasting Systems,
in which all processes receive a broadcast message at once.
A similar type of broadcast mechanism is used in the Broadcast Protocols of Emerson and Namjoshi \cite{Emerson-Namjoshi98}.
In our setting, this is similar to the case in which all
nodes share a common group (the underlying graph is a clique).
Ene and Muntean presented the $b\pi$-calculus~\cite{EM01},
an extension of the $\pi$-calculus~\cite{pi-calculus} with a broadcast
such that only nodes listening on the right channel
can receive emitted messages.
Wireless broadcast communication has been investigated in the context
of process calculi by Nanz and Hankin~\cite{NH06},
Singh, Ramakrishnan and Smolka~\cite{LISAT06,SCPJ08},
Lanese and Sangiorgi~\cite{LS10}, Godskesen~\cite{G07},
and Merro~\cite{Mer09}.
In particular Nanz and Hankin~\cite{NH06} consider
a graph representation of node localities to
determine the receivers
of a message, while Godskesen~\cite{G07} makes use of
a neighbour relation.
On the contrary, Lanese and Sangiorgi~\cite{LS10}
and Merro~\cite{Mer09} associate physical locations
to processes so that the receivers depend on the
location of the emitter and its transmission range.
As already mentioned, we have been directly inspired by the
$\omega$-calculus of Singh, Ramakrishnan and Smolka~\cite{LISAT06,SCPJ08}.
The $\omega$-calculus is based on the $\pi$-calculus.
The $\pi$-calculus \cite{pi-calculus} intermixes the communication and mobility of processes by expressing mobility as
change of interconnection structure among processes through communication.
In the $\omega$-calculus mobility of processes is abstracted from their communication actions, i.e., mobility is spontaneous and
it does not involve any communication.
In \cite{Concur09} the same authors define a constraint-based analysis for configurations with fixed topologies
and a fixed number of nodes. The authors also mention that checking reachability of a configuration from an initial one is decidable for
the fragment without restriction. This property is an immediate consequence of the fact that there is no dynamic generation or deletion
of processes (i.e. it boils down to a finite-state reachability problem).
The symbolic approach in \cite{Concur09} seems to improve verification results obtained with more standard model checking techniques.
For instance, in \cite{FVM07} model checking is used for automatic verification of finite-state and timed models of  Ad Hoc Networks.
In these works the number of nodes in the initial configurations is known and fixed a priori.
In order to detect protocol vulnerabilities tools like Uppaal are executed on all possible topologies (modulo symmetries)
for a given number of nodes.
In \cite{SWJ08} Saksena et al. define a symbolic procedure based on graph-transformations to analyze routing protocol for Ad Hoc Networks.
The symbolic representation is based on upward closed sets of graphs ordered w.r.t. subgraph inclusion.
Their procedure is not guaranteed to terminate. In our paper we consider a non trivial class of graphs (bounded path configurations) for which
backward analysis with a similar symbolic representation (upward closure of graphs w.r.t.  induced subgraph ordering)
is guaranteed to terminate for finite-state descriptions of individual nodes.

\paragraph{Structure of the paper}

In Section~\ref{sec:definitions} 
we formally introduce our model for ad hoc network protocols,
we define the parametric version of the control state reachability
problem,
and we recall the result from~\cite{delzanno-parameterized-10}, i.e. that control state reachability is
undecidable if we do not impose any restriction on the class
of possible topologies, while it turns out to be decidable when
restricting to bounded path topologies.
In Sections~\ref{sec:diameter} and~\ref{sec:cliques} 
we consider other restricted classes
that include clique graphs: bounded diameter and bounded path
on the maximal clique graph, respectively.
For these classes we report the results proved in~\cite{delzanno-fossacs11}: control state reachability is undecidable
when restricted to graphs with a bounded diameter (but it turns out
to be decidable if we additionally assume bounded degree),
while for the class of graphs having a corresponding
maximal clique graph with bounded path, the problem is decidable.
Section~\ref{conclusions} contains concluding remarks
and directions for future work.

\renewcommand{\gamma}{G}

\section{Ad Hoc Network Protocols}
\label{sec:definitions}
\subsection{Preliminaries on Graphs}

In this section we assume that $Q$ is a finite set of elements. A {\em $Q$-labeled undirected graph} (shortly $Q$-graph or graph) is a tuple $\gamma=(V,E,L)$, where $V$ is a finite set of {\em vertices} (sometimes called {\em nodes}), and $E\subseteq V\times V$ is a finite set of {\em edges}, and $L: V \rightarrow Q$ is a labeling function.
We consider here undirected graphs, i.e., such that $\tuple{u,v} \in E$ iff $\tuple{v,u} \in E$.
We denote by ${\cal G}_Q$ the set of $Q$-graphs. For an edge $\tuple{u,v} \in E$, $u$ and $v$ are called its {\em endpoints} and we say that $u$ and $v$ are adjacent vertices. For a node $u$ we call {\em vicinity} the set of its adjacent nodes (neighbors).
%
%
%
%
Given a vertex $v \in V$, the {\em degree} of $v$ is the size of the set $\set{u \in V \mid \tuple{v,u} \in E }$. The degree of a graph is the maximum degree of its vertices.
We will sometimes denote $\state(\gamma)$ the set $\state(V)$ (which is a subset of $Q$).
A {\em path} $\pi$ in a graph is a finite sequence $v_1,v_2,\ldots,v_m$ of vertices such that for $1 \leq i \leq m-1$, $\tuple{v_i,v_{i+1}} \in E$ and the integer $m-1$ (i.e. its number of edges) is called the length of the path $\pi$, denoted by $|\pi|$. A path $\pi=v_1,\ldots,v_m$ is simple if for all $1 \leq i,j \leq m$ with $i \neq j$, $v_i \neq v_j$, in other words each vertex of the graph occurs at most once in $\pi$. A {\em cycle} is a path $\pi=v_1,\ldots,v_m$ such that $v_1=v_m$.
%
%
%
%
A graph $\gamma=\tuple{V,E,L}$ is {\em connected} if for all $u,v \in V$ with $u \neq v$, there exists a path from $u$ to $v$ in $G$.
%
%
%
A {\em clique} in an undirected graph $\gamma=\tuple{V,E,L}$ is a subset $C \subseteq V$ of vertices, such that for every $u,v \in C$ with $u \neq v$, $\tuple{u,v} \in E$. A clique $C$ is said to be {\em maximal} if there exists
no vertex $u \in V \setminus C$ such that $C \cup \set{u}$ is a clique.
If the entire set of nodes $V$ is a clique, we say that $G$ is a clique graph.
%
%
%
A {\em bipartite $Q$-graph} is a tuple $\tuple{V_1,V_2,E,L}$ such that $\tuple{V_1 \cup V_2, E, L}$ is a $Q$-graph, $V_1 \cap V_2 = \emptyset$ and $E \subseteq (V_1 \times V_2) \cup (V_2 \times V_1)$.

%
%
The {\em diameter} of a graph $\gamma=\tuple{V,E,L}$ is the length of the {\em longest shortest simple path} between any two vertices of $G$.
Hence, the diameter of a clique graph is always one.
We also need to define some graph orderings.
Given two graphs $G=\tuple{V,E,L}$ and $G'=\tuple{V',E',L'}$,
$G$ is in the \emph{subgraph} relation with $G'$, written $G\preceq_s G'$, whenever there exists an injective function $f:V\rightarrow V'$ such that, for every $v,v'\in V$, if $\tuple{v,v'}\in E$, then $\tuple{f(v),f(v')}\in E'$ and for every $v \in V$, $L(v)=L'(f(v))$.
Furthermore, $G$ is in the \emph{induced subgraph} relation with $G'$, written $G\preceq_i G'$, whenever there
exists an injective function $f:V\rightarrow V'$ such that, for every $v,v'\in V$,  $\tuple{v,v'}\in E$ if and only if
$\tuple{f(v),f(v')}\in E'$ and for every $v \in V$, $L(v)=L'(f(v))$.
As an example, a path with three nodes  is a subgraph, but not an induced subgraph, of a ring of the same order.
Finally, we recall the notion of {\em well-quasi-ordering} (wqo for short).
A quasi order $(A,\leq)$ is a wqo if for every infinite sequence of elements
$a_1,a_2,\ldots,a_i,\ldots$ in $A$, there exist two indices $i<j$ s.t. $a_i\leq a_j$.
Examples of wqo's are the sub-multiset relation, and both the subgraph and the induced subgraph relation over
graphs with simple paths of bounded length \cite{Ding90}.

\subsection{Ad Hoc Networks}
\label{sec:sbp}\label{sec:model}
In our model of ad hoc networks a configuration is simply a graph
and we assume that each node of the graph is a process that runs a
common predefined protocol.
A protocol is defined by a communicating automaton with a finite set $Q$ of control states.
Communication is achieved via selective broadcast.
The effect of a broadcast is in fact local
to the vicinity of the sender. The initial configuration is any graph in which all the nodes are in an initial control state. Remark that even if $Q$ is finite, there are infinitely many possible initial configurations. We next formalize the above intuition.
\smallskip\\
{\em Individual Behavior}
 The protocol run by each node is defined via a process $\pdef=\tuple{Q,\Sigma,R,Q_0}$, where $Q$ is a finite set of control states, $\Sigma$ is a finite alphabet, $R \subseteq Q \times (\set{\mytau} \cup \set{\broadcast{a}, \receive{a} \mid a  \in \Sigma}) \times Q$ is the transition relation, and $Q_0 \subseteq Q$ is a set of initial control states.
The label $\mytau$ represents the capability of performing an internal action,
and the label $\broadcast{a}$ ($\receive{a}$) represents the capability of broadcasting (receiving) a message $a\in\Sigma$.
\smallskip\\
{\em  Network Semantics}
An AHN associated to $\pdef=\tuple{Q,\Sigma,R,Q_0}$ is defined via a transition system $\ahn_\pdef=\tuple{\confs,\Rightarrow,\confs_0}$,
where $\confs={\cal G}_Q$ (undirected graphs with labels in $Q$) is the set of configurations,
$\confs_0={\cal G}_{Q_0}$ (undirected graphs with labels in $Q_0$) is the subset of initial configurations,
and $\Rightarrow \subseteq \confs \times \confs$ is the transition relation defined next.
For $q \in Q$ and $a \in \Sigma$, we define the set $R_a(q)=\set{q' \in Q \mid \tuple{q,\receive{a},q'} \in R}$
that contains states that can be reached from the state $q$ upon reception of message $a$.
For $\gamma=\tuple{V,E,L}$ and $\gamma'=\tuple{V',E',L'}$, $\gamma \Rightarrow \gamma'$ holds iff  
$G$ and $G'$ have the same underlying structure, i.e., $V=V'$ and $E=E'$, and one of the following 
conditions on $L$ and $L'$ holds:
\begin{itemize}
\item  $\exists v\in V$ s.t.
$(L(v),\tau,L'(v)) \in R$, and $L(u)=L'(u)$ for all $u$ in $V \setminus \set{v}$;
\item  $\exists v \in V$ s.t. $(L(v),\broadcast{a},L'(v))\in R$  and for every $u \in V \setminus \set{v}$ 
\begin{itemize}
\item  if $\tuple{v,u} \in E$ and $R_a(L(u))\neq \emptyset$ (reception of $a$ in $u$  is enabled), then $L'(u) \in R_a(L(u))$.
\item  $L(u)=L'(u)$, otherwise.
\end{itemize}
\end{itemize}
An execution is a sequence $\gamma_0\gamma_1\ldots$ such that $\gamma_0 \in {\cal G}_{Q_0}$
and $\gamma_i\Rightarrow\gamma_{i+1}$ for $i\geq 0$.
We use $\Rightarrow^\ast$ to denote the reflexive and transitive closure of $\Rightarrow$.

Observe that a broadcast message $a$ sent by $v$ is delivered only to the subset of neighbors
interested in it. Such a neighbor $u$ updates its state with a new state taken from $R(L(u))$.
All the other nodes (including neighbors not interested in $a$) simply ignore the message.
Also notice that  the topology is static, i.e., the set of nodes and edges remain unchanged during a run.


Finally, for a set of $Q$-graphs $\Topology \subseteq {\cal G}_Q$, the AHN $A_{\pdef}^{\Topology}$ restricted to $\Topology$ is defined by the transition system $\tuple{\confs \cap \Topology,\Rightarrow_\Topology,\confs_0 \cap \Topology}$ where the relation $\Rightarrow_\Topology$ is the restriction of $\Rightarrow$ to $(\confs \cap \Topology) \times (\confs \cap \Topology)$.

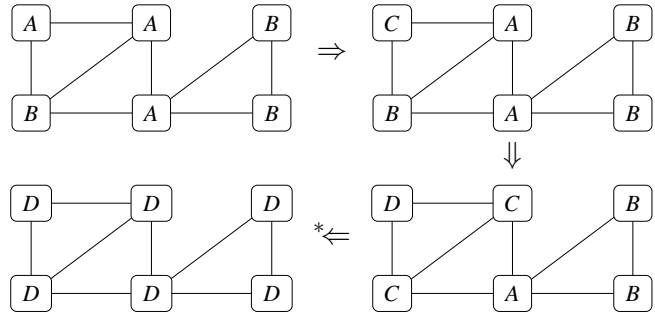
\begin{figure*}[t]
\begin{center}
\scalebox{0.8}{
\begin{picture}(100,45)(-20,-35)
  \gasset{Nadjust=w,Nadjustdist=2,Nh=6,Nmr=1,AHnb=0}
  \node[fillcolor=White](A)(-20,10){$A$}
  \node[fillcolor=White](B)(0,10){$A$}
  \node[fillcolor=White](C)(20,10){$B$}
  \node[fillcolor=White](D)(-20,-5){$B$}
  \node[fillcolor=White](E)(0,-5){$A$}
  \node[fillcolor=White](F)(20,-5){$B$}
  \drawedge[linecolor=Black](A,B){}
  \drawedge[linecolor=Black](A,D){}
  \drawedge[linecolor=Black](B,E){}
  \drawedge[linecolor=Black](B,D){}
  \drawedge[linecolor=Black](E,D){}
  \drawedge[linecolor=Black](C,E){}
  \drawedge[linecolor=Black](E,F){}
  \drawedge[linecolor=Black](C,F){}

\node[linecolor=White](Lab1)(30,5){\Large $\Rightarrow$}

  \node[fillcolor=White](A)(40,10){$C$}
  \node[fillcolor=White](B)(60,10){$A$}
  \node[fillcolor=White](C)(80,10){$B$}
  \node[fillcolor=White](D)(40,-5){$B$}
  \node[fillcolor=White](E)(60,-5){$A$}
  \node[fillcolor=White](F)(80,-5){$B$}
  \drawedge[linecolor=Black](A,B){}
  \drawedge[linecolor=Black](A,D){}
  \drawedge[linecolor=Black](B,E){}
  \drawedge[linecolor=Black](B,D){}
  \drawedge[linecolor=Black](E,D){}
  \drawedge[linecolor=Black](C,E){}
  \drawedge[linecolor=Black](E,F){}
  \drawedge[linecolor=Black](C,F){}

\node[linecolor=White](Lab2)(60,-12){\Large $\Downarrow$}

\node[fillcolor=White](A)(40,-20){$D$}
  \node[fillcolor=White](B)(60,-20){$C$}
  \node[fillcolor=White](C)(80,-20){$B$}
  \node[fillcolor=White](D)(40,-35){$C$}
  \node[fillcolor=White](E)(60,-35){$A$}
  \node[fillcolor=White](F)(80,-35){$B$}
  \drawedge[linecolor=Black](A,B){}
  \drawedge[linecolor=Black](A,D){}
  \drawedge[linecolor=Black](B,E){}
  \drawedge[linecolor=Black](B,D){}
  \drawedge[linecolor=Black](E,D){}
  \drawedge[linecolor=Black](C,E){}
  \drawedge[linecolor=Black](E,F){}
  \drawedge[linecolor=Black](C,F){}

\node[linecolor=White](Lab3)(30,-25){\Large $^\ast \hspace{-0.4em}\Leftarrow$}

  \node[fillcolor=White](A)(-20,-20){$D$}
  \node[fillcolor=White](B)(0,-20){$D$}
  \node[fillcolor=White](C)(20,-20){$D$}
  \node[fillcolor=White](D)(-20,-35){$D$}
  \node[fillcolor=White](E)(0,-35){$D$}
  \node[fillcolor=White](F)(20,-35){$D$}
  \drawedge[linecolor=Black](A,B){}
  \drawedge[linecolor=Black](A,D){}
  \drawedge[linecolor=Black](B,E){}
  \drawedge[linecolor=Black](B,D){}
  \drawedge[linecolor=Black](E,D){}
  \drawedge[linecolor=Black](C,E){}
  \drawedge[linecolor=Black](E,F){}
  \drawedge[linecolor=Black](C,F){}
\end{picture}
}
\end{center}
\caption{Example of execution}
\label{execution}
\end{figure*}
\subsection{Example of Ad Hoc Network Protocol}
\label{exampleAHN}
As an example of an ad hoc network protocol and of its semantics, consider a protocol consisting of  the following rules:
$(A,\tau,C)$,  $(C,\broadcast{m},D)$,  $(B,\receive{m},C)$,  and $(A,\receive{m},C)$.
As shown in Fig. \ref{execution}, starting from a configuration with only $A$ and  $B$ nodes, an $A$ node first moves to $C$ and then send $m$ to his/her neighbors. In turn, they forward the message $m$ to their neighbors, and so on.

\subsection{Decision problem}
We define the decision problem of {\em control state reachability} (\CSRP) as follows:
\begin{description}
\item[Input:] A process $\pdef=\tuple{Q,\Sigma,R,Q_0}$ with $\ahn_\pdef=\tuple{\confs,\Rightarrow,\confs_0}$ and a control state $q \in Q$;
\item[Output:] Yes, if there exists
$\gamma \in \confs_0$ and $\gamma' \in \confs$ such that $q \in \state(\gamma')$ and
$\gamma \Rightarrow^\ast \gamma'$, no otherwise.
\end{description}
Control state reachability is strictly related to parameterized verification of safety properties. 
The input control state $q$ can in fact be seen as an error state for the execution of the protocol in some node of the network.
If the answer to \CSRP~is yes, then there exists a sufficient number of processes, all executing the same protocol, 
and an initial topology from which we can generate a configuration in which the error is exposed.
%
Under this perspective, \CSRP~ can be viewed as instance of a parameterized verification problem.

In~\cite{delzanno-parameterized-10} we have proved that
\CSRP~is undecidable. 
The proof is by reduction from the halting problem for two-counter
Minsky machines. A Minsky machine manipulates two integer variables $c_1$ and $c_2$, which are called counters, and it is composed of a finite set of instructions. Each of the instuction is either of the form (1) $L: c_i :=c_i+1; \mathtt{~goto~} L'$ or (2) $L: \mathtt{if~} c_i=0 \mathtt{~then~goto~} L' \mathtt{~else~} c_i :=c_i-1; \mathtt{~goto~} L''$ where $i \in \set{1,2}$ and $L,L',L''$ are labels preceding each instruction. Furthermore there is a special label $L_F$ from which nothing can be done. The halting problem consists then in deciding whether or not the execution that starts from $L_0$ with counters equal to $0$ reaches $L_F$.

The intuition behind the reduction is as follows.
In a first phase we exploit an exploration protocol
to impose a logical topology on top of the
actual physical node connections. This logical topology
is composed by a control node which is
connected to two distinct lists of nodes used to simulate the content of the counters.
 Each node in the list associated to counter $c_i$ is either in state $Z_i$ or $NZ_i$. The current value of the counter $c_i$ equals the number of $NZ_i$ nodes in the list.
The length of each list is guessed non-deterministically during the execution of the first phase (i.e. before starting the simulation) and it corresponds to the maximum value store in a counter for the simulation to succeed.
Initially, all nodes must encode zero (state $Z_i$).

In the second phase the control node starts the simulation of the instructions.
It operates by sending requests that are propagated back and forth a list by using broadcast 
sent by a node to its (unique) single-hop successor/predecessor node.
The effect of these requests is to change the state of one node in zero state $Z_i$ to the non-zero state $NZ_{i}$ in case of increment, or the
vice versa in the case of decrement. The test-for-zero instruction on
the counter $c_{i}$ is simply simulated by checking whether there are no  nodes in the zero state $Z_{i}$ in the i-th list.

\subsection{Configurations with Bounded Path}

In~\cite{delzanno-parameterized-10} we have proved that \CSRP~turns
out to be decidable if we restrict the possible topologies to 
the class of graphs whose path is bounded by $k$ (for a fixed $k >0$). 
The proof is based on the theory of Well Structured Transition Systems \cite{lics96,ic2000,FinSch01} (WSTS). A WSTS is a transition
system equipped with a well-quasi ordering on states and a monotonicity
property: if a configuration $c_1$ smaller than a configuration $c_2$ has
a transition to a configuration $c'_1$, then also $c_2$ has a transition to a 
configuration $c'_2$ which is greater than $c'_1$. Coverability turns out to be decidable in WSTSs by using backward analysis, if it is possible to compute the predecessors of a given state.

In~\cite{delzanno-parameterized-10} we have observed that ad hoc network
protocols are monotonic with respect to the induced subgraph ordering relation, while this is not the case for the subgraph ordering relation. This is already an interesting observation that distinguishes selective broadcast from point-to-point communication, which is monotonic with respect to the usual subgraph ordering. The proof of decidability
is completed by defining how to compute the predecessors,
and by observing that the induced subgraph ordering is a wqo for the class of graphs for which the length of simple paths is bounded by a constant (i.e. bounded path graphs). This result is known as Ding's Theorem \cite{Ding90}.

\section{Configurations with Bounded Diameter}
\label{sec:diameter}

As mentioned in the introduction,
restricting protocol analysis to configurations 
with bounded path 
seems to have a limited application in a communication model with selective broadcast.
For these reasons, in~\cite{delzanno-fossacs11} we have investige \CSRP~for restricted classes of graphs that at least include the class of clique graphs.
\begin{figure}[t]
\begin{center}
\scalebox{0.9}{
\begin{picture}(40,40)(-20,-20)
\gasset{Nadjust=w,Nadjustdist=2,Nh=6,Nmr=1,AHnb=0,ATnb=0}
  \node[fillcolor=White](C)(0,0){$L_0$}
\node[fillcolor=White](1)(-60,20){$firstZ_1$}
  \node[fillcolor=White](2)(-40,20){$Z_1$}
  \node[fillcolor=White](3)(-20,20){$Z_1$}
\node[fillcolor=White](4)(20,20){$Z_1$}
\node[fillcolor=White](5)(40,20){$Z_1$}
\node[fillcolor=White](6)(60,20){$Z_1$}
\node[linecolor=white](a)(0,20){\Large{$\ldots$}}

\node[fillcolor=White](7)(-60,-20){$firstZ_2$}
  \node[fillcolor=White](8)(-40,-20){$Z_2$}
  \node[fillcolor=White](9)(-20,-20){$Z_2$}
\node[fillcolor=White](10)(20,-20){$Z_2$}
\node[fillcolor=White](11)(40,-20){$Z_2$}
\node[fillcolor=White](12)(60,-20){$Z_2$}
\node[linecolor=white](b)(0,-20){\Large{$\ldots$}}

  \drawedge[linewidth=0.5](C,1){}
  \drawedge[linewidth=0.5](C,2){}
  \drawedge[linewidth=0.5](C,3){}
\drawedge[linewidth=0.5](C,4){}
\drawedge[linewidth=0.5](C,5){}
\drawedge[linewidth=0.5](C,6){}
	\drawedge[linewidth=0.5](1,2){}
\drawedge[linewidth=0.5](2,3){}
\drawedge[linewidth=0.5](3,a){}
\drawedge[linewidth=0.5](a,4){}
\drawedge[linewidth=0.5](4,5){}
\drawedge[linewidth=0.5](5,6){}
\drawedge[linewidth=0.5](C,7){}
  \drawedge[linewidth=0.5](C,8){}
  \drawedge[linewidth=0.5](C,9){}
\drawedge[linewidth=0.5](C,10){}
\drawedge[linewidth=0.5](C,11){}
\drawedge[linewidth=0.5](C,12){}
	\drawedge[linewidth=0.5](7,8){}
\drawedge[linewidth=0.5](8,9){}
\drawedge[linewidth=0.5](9,b){}
\drawedge[linewidth=0.5](b,10){}
\drawedge[linewidth=0.5](10,11){}
\drawedge[linewidth=0.5](11,12){}

\end{picture}}
\end{center}
\caption{Butterfly-shaped induced subgraph needed to simulate a Minsky machine.}
\label{butterfly}
\end{figure}
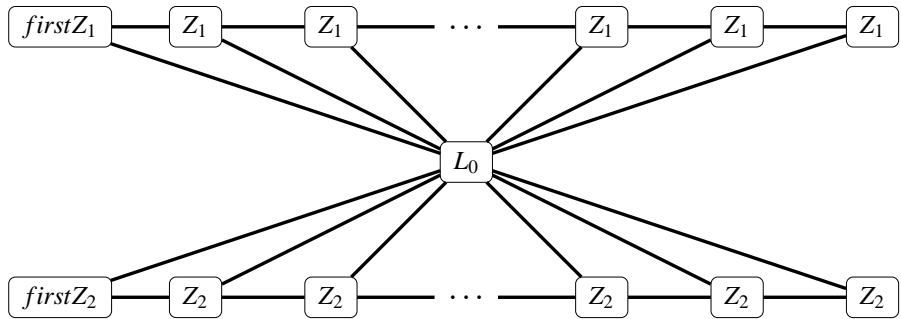
%
The first class we have consider is that of graphs with bounded diameter.
  Fixed $k>0$, a graph $\gamma$ has a $k$-bounded diameter if and only if its diameter is smaller than or equal to $k$.
Observe that for every $k>0$, clique graphs belong to the class of graphs with a diameter bounded by $k$.  Furthermore, given $k>0$ the class 
of graphs with path bounded by $k$
is included in the class of graphs with a diameter bounded by $k$.
Graphs with $k$-bounded diameter coincide with the so called $k$-clusters used in partitioning algorithm for ad hoc networks \cite{Clusters}. Thus, this class is of particular relevance for the analysis of selective broadcast communication. Intuitively, the diameter corresponds to the minimal number of broadcasts (hops) needed to send a message to all nodes connected by a path with the sender.

The \CSRP~problem restricted to configurations with $k$-bounded diameter turns out to be undecidable for $k>1$.
The proof is similar to the proof of undecidability for the
general case reported in~\cite{delzanno-parameterized-10}:
by reduction from the halting problem
for two-counter Minsky machines.

The main difference is that the logical topology to be imposed
in the first phase of the simulation of the Minsky machines should
be with bounded diameter (namely, diameter 2). The 
topology that we have considered 
is a sort of 
butterfly (see Figure \ref{butterfly}) consisting of two lists (to represent the counters) and in which all nodes in the lists are connected to a monitor node (to represent the program counter).
The second phase of the simulation, i.e. the actual execution of the
instructions, proceeds similarly to the protocol described above.
The unique difference is that we use a distinct $firstZ_{i}$ node
to distinguish the initial node of each list (this is needed as
now all the list nodes are connected to the program counter node).

Note that if we restrict our attention to graphs with a diameter bounded by $1$, the above encoding does not work anymore. The class of
graphs with diameter 1 corresponds to the set of clique graphs
and, as said above, {\CSRP} turns out to be decidable when restricting
to clique topologies.

\subsubsection*{Bounded diameter and bounded degree.}
%
From a non trivial result on bounded diameter graphs \cite{hoffman-moore-60},  we have obtained in~\cite{delzanno-fossacs11} an interesting decidable subclass.
Indeed, in \cite{hoffman-moore-60} the authors show that, given two integers $k,d >0$, the number of graphs whose diameter is smaller than $k$ and whose degree is smaller than $d$ is finite.
The Moore bound $M(k,d)=(k(k-1)^d-2)/(k-2)$ is an upper bound for the size of the largest undirected graph in such a class. 
It follows that, for $k,d>0$, and an ad hoc protocol with $n$ states,
if we restrict to configurations with a diameter bounded by $k$ and a degree bounded by $d$, the state space is bounded by $n^{M(k,d)}$,
thus it is polynomial in the size of the protocol. Consequently we 
can conclude that \CSRP~restricted to configurations with
$k$-bounded diameter and $d$-bounded degree is in \textsc{Pspace}.

%
%
%
%
%

%
%


\newcommand{\BPC}{BPC}
\section{Maximal Clique Graphs with Bounded Paths}
\label{sec:cliques}

In this section we describe classes of graphs 
that strictly increases both the classes of clique graphs and
the classes of bounded path graphs, for which we have proved
in~\cite{delzanno-fossacs11} that \CSRP~is decidable.
We have called these classes of graphs $\BPC_n$ ($n$-Bounded Path maximal Cliques graphs). Namely, for $n>0$
$\BPC_n$ contains both
$n$-bounded path graphs and any clique graph, while being strictly contained in
the class of graphs with $2n$-bounded diameter.
These classes are defined on top of the notion of {\em maximal clique graphs} associated to a configuration.
\begin{definition}
Given a connected undirected graph $G=\tuple{V,E,L}$ and $\bullet\not\in L(V)$,
the {\em maximal clique graph}  $K_G$ is the bipartite graph  $\tuple{X,W,E',L'}$ in which
\begin{itemize}
\item $X=V$;
\item $W\subseteq 2^V$ is the set of maximal cliques of $G$;
\item For $v\in V,w\in W$, $\tuple{v,w}\in E'$ iff $v\in w$;
\item $L'(v)=L(v)$  for $v\in V$, and $L'(w)=\bullet$ for $w\in W$.
\end{itemize}
\end{definition}
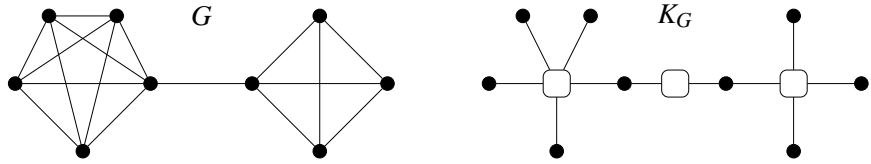
\begin{figure}[t]
\begin{center}
\scalebox{0.9}{
\begin{picture}(100,15)(10,20)
\gasset{fillcolor=black,Nadjustdist=2,Nh=2, Nw=2,Nmr=1,AHnb=0,ATnb=0}
  \node(n1)(0,35){}
  \node(n2)(10,35){}
  \node(n3)(-5,25){}
  \node(n4)(15,25){}
  \node(n5)(5,15){}
  \drawedge(n1,n2){}
  \drawedge(n1,n3){}
  \drawedge(n1,n4){}
  \drawedge(n1,n5){}
  \drawedge(n2,n3){}
  \drawedge(n2,n4){}
  \drawedge(n2,n5){}
  \drawedge(n3,n4){}
  \drawedge(n3,n5){}
  \drawedge(n4,n5){}

  \node(n6)(40,35){}

  \node(n7)(30,25){}
  \drawedge(n4,n7){}
  \node(n8)(40,15){}
  \node(n9)(50,25){}
   \drawedge(n6,n7){}
  \drawedge(n6,n8){}
  \drawedge(n6,n9){}
  \drawedge(n7,n8){}
  \drawedge(n7,n9){}
  \drawedge(n8,n9){}

  \node[fillcolor=white,linecolor=white,Nh=0,Nw=0](l1)(22.5,35){\large$G$}

  \node(n1b)(70,35){}
  \node(n2b)(80,35){}
  \node(n3b)(65,25){}
  \node(n4b)(85,25){}
  \node(n5b)(75,15){}

  \node[fillcolor=white,Nh=4,Nw=4](n1c)(75,25){}
 \drawedge(n1b,n1c){}
    \drawedge(n2b,n1c){}

  \drawedge(n3b,n1c){}
  \drawedge(n1c,n5b){}
  \drawedge(n4b,n1c){}
  \node(n6b)(110,35){}

  \node(n7b)(100,25){}

  \node(n8b)(110,15){}
  \node(n9b)(120,25){}

  \node[fillcolor=white,linecolor=white,Nh=0,Nw=0](l2)(92.5,35){\large$K_G$}
  \node[fillcolor=white,Nh=4,Nw=4](n2c)(92.5,25){}
  \drawedge(n4b,n2c){}
\drawedge(n2c,n7b){}

\node[fillcolor=white,Nh=4,Nw=4](n3c)(110,25){}
\drawedge(n6b,n3c){}
    \drawedge(n3c,n8b){}
  \drawedge(n7b,n3c){}
  \drawedge(n3c,n9b){}

  \end{picture}}
  \end{center}
\caption{A graph $G$ and its associated clique graph $K_G$.}
\label{clique-graph}
\end{figure}
Note that for each connected graph $G$ there exists a unique maximal clique graph $K_G$. An example of construction is given by Figure \ref{clique-graph}. One can also easily prove that if $G$ is a clique graph then in $K_G$ there is no path of length strictly greater than $3$. Furthermore, from the maximality of the cliques in $W$ if two nodes $v_1,v_2\in V$ are connected both to $w_1$ and $w_2\in W$, then $w_1$ and $w_2$ are distinct cliques.
We use the notation $\inclique{v_1}{v_2}{w}$ to denote that $v_1,v_2$ belong to the same clique $w$.

%
\begin{definition}
For $n\geq 1$, the class $\BPC_n$ consists of the set of configurations whose associate
maximal clique graph has $n$-bounded paths (i.e. the length of the simple paths of $K_G$ is at most $n$).
\end{definition}

The proof of decidability of {\CSRP} for $\BPC_{n}$ graphs
is based on an ordering defined on maximal clique graphs
that corresponds to the induced subgraph ordering defined
on the corresponding graphs. Such a new ordering is defined 
as follows.
\begin{definition}
Assume $G_1=\tuple{V_1,E_1,L_1}$ with $K_{G_1}=\tuple{X_1,W_1,E_1',L_1'}$,  and $G_2=\tuple{V_2,E_2,L_2}$ with $K_{G_2}=\tuple{X_2,W_2,E_2',L_2'}$ with $G_1$ and $G_2$ both connected graphs.
Then, $G_1\sqsubseteq G_2$ iff there exist two injective functions $f:X_1\rightarrow X_2$ and $g:W_1\rightarrow W_2$,
such that
\begin{description}

\item[(i)] for every $v\in X_1$, and $C\in W_1$, $v\in C$ iff  $f(v)\in g(C)$;

\item[(ii)] for every $v_1,v_2\in X_1$, and $C\in W_2$,
      if $\inclique{f(v_1)}{f(v_2)}{C}$, then there exists $C'\in W_1$ s.t. $\inclique{f(v_1)}{f(v_2)}{g(C')}$;

\item[(iii)] for every $v\in X_1$, $L'_1(v)=L'_2(f(v))$;
\item[(iv)] for every $C\in W_1$, $L'_1(C)=L'_2(g(C))$.
\end{description}
\end{definition}
The first condition ensures that (dis)connected nodes remain (dis)connected inside the image of $g$.
Indeed, from point (i) it follows that, for every $v_1,v_2\in X_1$, and $C\in W_1$,
 $\inclique{v_1}{v_2}{C}$ iff  $\inclique{f(v_1)}{f(v_2)}{g(C)}$.
The second condition ensures that disconnected nodes remain disconnected outside the image of $g$.

By condition (i) in the definition of $\sqsubseteq$, we also have that  $G_1\sqsubseteq G_2$ (via $f$ and $g$)
implies that  $K_{G_1}$ is in the induced subgraph relation with $K_{G_2}$ (via $f\cup g$). The relation between this new relation
and the induced subgraph ordering is even stronger, in fact we 
have proved in~\cite{delzanno-fossacs11} that the two
coincide:
$G_1\sqsubseteq G_2$ iff $G_1$ is  an induced subgraph of $G_2$.

The main theorem in~\cite{delzanno-fossacs11} states that
for any $n\geq 1$, $(\BPC_n,\sqsubseteq)$ is a well-quasi ordering.
In the light of the correspondance result between $\sqsubseteq$
and the induced subgraph ordering, and the monotonicity of
ad hoc network protocol 
with respect to the induced subgraph ordering relation
(and the computability of the predecessors) discussed the previous section, we have been able to prove in~\cite{delzanno-fossacs11} 
the decidability of {\CSRP} for topologies restricted to graphs in $\BPC_n$ (for a fixed $n >0 $).

In~\cite{delzanno-fossacs11}
we have investigated also the complexity of the decision procedure
for \CSRP\ restricted to topologies in $\BPC_n$.
We have found that this problem
is  not primitive recursive.
The proof is by reduction from the coverability
problem for reset nets, which is known to be an
Ackermann-hard problem~\cite{Schnoebelen2010}.

\section{Conclusions}\label{conclusions}

In this paper we have reported the main result that we have
recently proved in~\cite{delzanno-parameterized-10,delzanno-fossacs11}
about the decidability and complexity boundaries for
the decidability and the complexity
of the parametric verification of safety properties
in ad hoc networks. Namely, given an ad hoc network
protocol expressed as a finite state communicating
automaton, we are interested in checking the existence
of an initial network configuration that can generate
a computation leading to a configuration in which at
least one node is in a given (error) state.

The problem is undecidable if no restrictions are imposed
to the possible initial configurations, but it turns out
to be decidable for interesting classes of graphs
in which the corresponding maximal cliques are connected by paths of bounded length. These graphs include both cliques and bounded path graphs.
The problem returns to be undecidable for bounded diameter
graphs. 

As a future work, we plan to study decidability and complexity
issues in presence of communication and node failures.
In particular, an interesting case of communication
failure in the context of ad hoc networks is due to
{\em conflicts} deriving form the contemporaneous emission
of signals from two distinct nodes that share some
neighbors. We plan to move to a truly concurrent semantics
for ad hoc network protocols in order to faithfully represent
this specific phenomenon.

\bibliographystyle{eptcs}

%

\end{document}